\documentclass[english]{article}
\usepackage[T1]{fontenc}
\usepackage[latin9]{inputenc}
\usepackage{geometry}
\usepackage{float}
\usepackage{babel}
\usepackage{booktabs}
\usepackage{amsmath}
\usepackage{appendix}
\usepackage[unicode=true,
 bookmarks=false,
 breaklinks=false,pdfborder={0 0 1},backref=section,colorlinks=false]
 {hyperref}
\usepackage{breakurl}

\makeatletter

\usepackage{fullpage}
\usepackage{changepage}
\usepackage{graphicx}
\usepackage{morefloats}
\usepackage{verbatim}
\usepackage{pdflscape}
\usepackage{ifthen}
\usepackage{verbatim}
\usepackage[justification=centering]{caption}
\usepackage{subcaption}
\usepackage[all]{hypcap}

\def\DHLhksqrt#1#2{%
\setbox0=\hbox{$#1\sqrt{#2\,}$}\dimen0=\ht0
\advance\dimen0-0.2\ht0
\setbox2=\hbox{\vrule height\ht0 depth -\dimen0}%
{\box0\lower0.4pt\box2}}

\usepackage{babel}

\makeatother

\begin{document}

\title{$J=0, J=J_{max},$ and Quadrupole Pairing}

\author{Larry Zamick and Daniel Hertz-Kintish\\
 \textit{Department of Physics and Astronomy, Rutgers University,
Piscataway, New Jersey 08854}}
\maketitle
\begin{abstract}
We consider 2 neutrons and 2 protons in the $g_{9/2}$ shell.Wave
functions and energy levels are obtained for various interactions.
The wave functions for states with total angular momentum I greater
or equal to 10 are not affected by what the pairing interacton (J=0
T=1) is.Other parts of the interaction are therefore of increased
importance. 
\end{abstract}

\section{Introduction }

In 1964, McCullen et al. {[}1{]} (MBZ) included detailed wavefunctions
in the $f_{7/2}$ shell in their paper and an accompanied technical
report. At that time the spectrum of the J=0 pairing Hamiltonian was
well studied. For the lowest state of $^{44}$Ti with this interaction
the coefficients are:
$$C(0,0)=0.8660, C(22) =.2152 ,C(44)= 0.2887 C(6,6) =0.3469.$$

However MBZ obtained quite different coefficients with $C(0,0)$ smaller
and $C(2,2)$ bigger. The wavefunction coefficients in an updated
version of MBZ by Escuderos et al. {[}2{]} are:
$$C(0,0) =0.7878 , C(2,2) = 0.5615, C(4,4) =0.2208, C(6,6)=0.1234.$$

Whereas the (J=0 T=1) pairing interaction focuses on particles of one
kind, with a system of both protons and neutrons one can have isospin
T=0 interactions and the ones with J=1$^{+}$ and J=J$_{max}(=7)$
lie low. 

For 2 protons and 2 neutrons in the $g_{9/2}$ shell, a remarkably
similar wavefuction structure was obtained by the Swedish group {[}3{]}.
The coefficients are:
$$C(0,0)=0.76, C(2,2)=0.57, C(4,4)=0.24 ,C(6,6)=0.13,\text{ and }C(8,8)= 0.14.$$

History repeats itself. They however developed
the conseqences of J$_{max}$ pairing more sharply than was done in
the past.

In a recent paper on maximum J pairing, Zamick and Escuderos {[}4{]}
made a comparison of spectra and wavefunctions of various schematic
interactions with those of a more realistic interaction. Specifically
they considered 2 proton holes and 2 neutron holes relative to doubly
magic $^{100}$Sn, i.e. $^{96}$Cd. These were single $j$ shell calculations
in the $g_{9/2}$ shell. In particular, in Table VI of {[}1{]}, the authors
compare overlaps of wavefunctions of two schematic interactions. We
use the notation $E(J)$ for an interaction in which all two-body matrix
elements are zero except the one with angular momentum J. In {[}1{]}
we have compared the overlaps of wavefunctions arising from matrix
diagonalizations with $E(9)$ and $E(0,9)=(E(0) + E(9))$ with wavefunctions
arising from the realistic interaction CCGI {[}2{]} represented by
10 matrix elements $V(J)=\langle(jj)^{I}|V|(jj)^{I}\rangle$, I=0,1,2,...,9.
These are respectively -2.3170, -1.4880, -0.6670, -0.4400, -0.1000,
-0.2710, 0.0660, -0.4040, 0.2100, and -1.4020.

It should be mentioned that in re. {[}1{]} Table VI, to an excellent
approximation, the $E(9)$ interaction for the wavefunctions of the lowest
states were proportional to unitary $9j$ symbols $\psi=N\langle(jj)^{9}(jj)^{9}|(jj)^{J_{p}}(jj)^{J_{n}}\rangle^{I}$
with $J_{p}$ and $J_{n}$ both even and $N$ aproximately $1/\sqrt{2}$.

\section{Overlaps}

In this section we will consider overlaps $\langle\psi$, $\psi_{CCGI}\rangle$.
Let us somewhat arbritarily say that anything greater than 0.9 is
a good overlap. When we use the $E(9)$ interaction, we get good overlaps
for the lowest energy states with $I=0,2,\text{ and }4$. We get bad overlaps
for $I= 6,8,\text{ and }10$, the values being 0.6795, 0.2375, and 0.6860 respectively. For
$I=12,14,\text{ and }16$, we again get good overlaps. When we use the $E(0,9)$
interaction we get still good overlaps for $I=1,2,3,\text{ and }4$ but in
addition we get good overlaps for $I=6\text{ and }8$, the latter two being 0.9361
and 0.9858. That is to say equal attraction in The J=0 and J=J$_{max}$
channels cures the problem that is present when the only interaction
is in the J=9 channel.

It should be noted however that when we switch from $E(9)$ to $E(0,9)$
there is no change in the overlaps for $I=10,12,14,\text{ and }16$. This is
easy to understand. For these high angular momentum states there cannot
be a pair of nucleons coupled to angular momentum zero, because the maximum angular momentum of the other two nucleons in the
$g_{9/2}$ shell is 9.

This leaves $I=10$ as a special state, still with a bad overlap of 0.6860.
Clearly, since J=0 pairing is out of the picture, this state will be
more sensitive to other parts of the interaction. We therefore consider $E(1,9) = [E(1) +E(9)], E(2,9) = 1/2 E(2) +E(9)\text{
as well as }E(1,2,9) E(1)+1/2E(2)+E(9)$. We take all the interactions to be
attractive (negative). We get the following overlaps for $I=10$:

\begin{table}[H]
\centering\protect\caption{$I=10$ Overlaps with Unitary $9j$ Coefficients}
\begin{tabular}{rr}
Interaction & Overlap\\
\toprule
$E(9)$& 0.6830\\

$E(2,9)$& 0.9004\\

$E(1,9)$& 0.9055\\

$E(1,2,9)$& 0.9659\\
\bottomrule
\end{tabular}
\end{table}

Note that $E(2,9)$ and $E(1,9)$, although quite different interactions,
give almost the same overlaps. This means it is dangerous to determine
the parameters of an interaction solely on the basis of overlaps.
We get the best overlap by lowering all 3 matrix elements, J=1, 2,
and 9. This agrees qualitatively with the CCGI interaction {[}2{]}.

Concerning the highest angular momentum state, I =16, this is a unique
state so it is not surprising that the overlap with any interaction
is one. What about $I=12$ and14? We not that for these not only does
$E(0)$ not enter but also $E(2)$ (2+9 =11). The fact that $E(0)$ and $E(2)$ do not
enter is clealy shown in Fig. 1 of the work of Qi {[}3{]}. Related
works by the' ``Swedish group'' are here also cited {[}3,4,5,6{]}
as well as the almost-Swedish group {[}7{]}.

Seniority arguments have been presented for the 8$^{+}$ state by
Fu, et al. {[}8,9{]} and the ``Swedish group''{[}4{]}. They note
this state is not well described as $[9\text{ }9]^{8}$, i.e. 2 pairs of
neutrons and protons each coupled to spin 9, but rather as seniority
2 state [8 0] and [0 8]. Here in the first term we have 2
protons coupled to 8$^{+}$ and 2 neutrons to 0$^{+}$, etc. For
CCGI, the pobability of these 2 conffigurations is 66\%.

\section{Overlaps of $E(9)$ with $U9j$}

In the following tables, we present overlaps of wavefuntions for the
$E(9)$ interaction and the column vectors of the $U9j$ interaction properly
normalized. Here $U9j=\langle (jj)^{9} (jj)^{J_{B}} | (jj)^{J_{P}}
(jj)^{J_{N}} \rangle^{I}$. We see that with $J_{B}=J_{max}=9$,
we get overlaps very close to one for the lowest states for all even
$I$. For $I=2$, there are very strong overlaps for the first two states with
associations $J_{B}=9$ and 7 respectively.

We give special attention to $I=4$. Beyond the lowest $I=4$ state there
are two nearly degenerate states at 3.028 MeV and 3.072 MeV. The overlaps
are 0.626255 and 0.786196 for the lowest state corresponding to $J_B=7$ and $J_B=5$,
respectively, and -0.779731 and 0.617937 for the upper state. However, we
can take linear combinations of the two eigenstates of the $E(9)$ interaction
$\alpha |1\rangle + \beta|2\rangle$ and $-\beta|1\rangle + \alpha |2\rangle$ such that
the overlaps are very close to one with the first state associated
with $J_B=7$ and the second with $J_{B}=5$. The coefficients
$\alpha$ and $\beta$ are obtained as follows:

Let $\vec{a}=|1\rangle$ be the first $E(9)$ eigenstate, $\vec{b}=|2\rangle$ be the second, and $\vec{U}$ be the U9j column vector for a given value of $J_B$. We use the Lagrange Multipliers method to maximize the function $f(\alpha,\beta)=(\alpha\vec{a}+\beta\vec{b})\cdot\vec{U}$ with the constraint $g(\alpha,\beta)=\alpha^2+\beta^2=1$. This method gives the extrema of $f$ with constraint $g$ as solutions to $\nabla f=\lambda\nabla g$ for some real constant $\lambda$. Using this method, we find

$$\alpha^2=\frac{m^2}{m^2+1},m\equiv\frac{\vec{a}\cdot\vec{U}}{\vec{b}\cdot\vec{U}}$$
and $\beta^2=1-\alpha^2$. Note that $m$ is simply the ratio of the overlaps.

Note that we assign isospin quantum numbers to the states in the appendix.
We can do this because the $E(9)$ interaction and indeed all the interactions
considered here are charge independent.

It should be noted that with the $E(0)$ interaction, the $T=0$ ground state
cannot be 100\% $(J_p, J_n)=(0,0)$ {[}2{]}. This is because
the unique $T=2$ states must have substational amount of this configuration.
Recent work by K. Neergaard shows that one can get improved $I=0^+$
wavefunctions in $^{48}$Cr, $^{88}$Ru, and $^{92}$Pd by admixing
75\% $(s,t) = (0,0)$ and 25\% $(4,0)$ where $s$ is seniority and $t$ reduced
isospin for $Sp(2j+1)$ {[}10{]}.

\section{Conclusions}

In this simplified model, we find a priori the surprising behavior
that overlaps of the $E(9)$ interaction with realistic interactions
exceed the 0.9 limit for low angular momenta $(I=0,2,4)$, are well
below the 0.9 limit for intermediate $I$ (6,8,10), but again exceed this
limit for large $I$ (12,14 and 16). Introducing $E(0,9)$ cures the problem
for $I=6$ and 8 but not for $I=10$. For $I=10$, the details of the J=2 matrix
element become super imprortant (quadrupole pairing), but J=2 does
not affect $I=12,14,$ and 16.

It should be cautioned that overlaps can be deceptive. For example,
for $I=0$, the overlap with $E(9)$ is 94.5\% {[}1{]}, but the spectrum
is such that the J=16$^{+}$ state is the ground state. It is 1.06
MeV below the lowest $I=0^{+}$ state. With $E(1,9), E(2,9)\text{ and }E(1,2,9)$,
we get a reversal with $I=16^{+}$ above $I=0^{+}$. The respecive
values are 0.920,3.332 and 4.947 MeV. We of course know that all even-even
nuclei have $I=0^{+}$ ground states. Indeed with the realistic CCGI
interaction{[}2{]}, the $I=16^{+}$ state is at an excitation energy
of 5.244 MeV (above the $I=0^{+}$ ground state){[}1{]}. 

What are the manifestations of J$_{max}$ pairing or more generally of
the proton-neutron interaction? This is most easily discussed by considering
a system of 2 protons and 2 neutrons described by a wavefunction
$\varSigma C(J_{p}J_{n}) [J_{p} J_{n}]$. This is
actually an old story. 

The values of $C(J,J)$ for various schematic interactions in the $g_{9/2}$
shell are here given as well as the $I=16^{+} - I=0^{+}$ splitting
(in square brackets). 

\begin{table}[H]
\centering \protect\caption{$C(J,J)$ for Various Interactions}
\begin{tabular}{rcrl}
Interaction&$C(J,J)$&$I$ Splitting\\
\toprule 
\addlinespace
E(0)& (0.8563, 0.1714, 0.2335, 0.2807, 0.3210)& {[}4.4000{]}& J=0 pairing\\
E(9) &(.6104, 0.7518, 0.2328, 0.0233, 0.0005) &{[}-1.0589{]} &J$_{max}$ pairing\\
E(1) &(-0.4675, 0.3836, 0.0725, 0.7174, 0.3836)& {[}3.6494{]}\\
E(0,9)& (0.8013, 0.4814, 0.2514, 0.1718, 0.1833)& {[}2.1678{]}\\
E(1,9) &(0.5202,0.7271,0.3724, 0.0831, -0.2335) &{[}0.9025{]}\\
E(2,9) &(0.3701,0.9602, 0.1077, -0.1585,-0.0722 ) &{[}3.3332{]}\\
E(1,2,9)& (0.3765,0.8787,0.2098,-0.0922,-0.1833 &{[}4.9467{]}\\
CCGI &(0.7725, 0.5289, 0.2915, 0.1704, 0.10210 &{[}5.2447{]}\\
\end{tabular}
\end{table}

Daniel Hertz-Kintish thanks the Rutgers Aresty Research Center for
Undergraduates for support during the 2014 summer session.

\appendix{

\renewcommand{\appendixname}{Overlaps for Even I and Odd $J_B$}
\appendixpage

Overlaps for even $I$ and odd $J_B$ ($T=0$):
\begin{table}[H]
\centering \protect\caption{$I=2$}
\begin{tabular}{r|rr}
E$\textbackslash J_B$ & 9 & 7\\
\toprule
1.0589 & 0.999959 & 0.000503096\\
3.0558 & -0.000175206 & 1.00003\\
4.058 & 0 & 0\\
5.0588 & 0 & 0\\
5.0588 & 0 & 0\\
5.0588 & 0 & 0\\
5.0588 & 0 & 0\\
6.0588 & 0 & 0\\
6.0588 & 0 & 0\\
6.0588 & 0 & 0\\
8.0588 & 0 & 0\\
8.0588 & 0 & 0\\
\end{tabular}
\end{table}

\begin{table}[H]
\centering \protect\caption{$I=4$}
\begin{tabular}{r|rrr}
E$\textbackslash J_B$ & 9 & 7 & 5\\
\toprule
1.0588 & 1.00003 & -0.000373778 & 0.00285081\\
3.028 & -0.000954468 & 0.626255 & 0.786196\\
3.0715 & -0.00105456 & -0.779731 & 0.617937\\
4.0308 & 0 & 0 & 0\\
4.0605 & 0 & 0 & 0\\
5.0588 & 0 & 0 & 0\\
5.0588 & 0 & 0 & 0\\
5.0588 & 0 & 0 & 0\\
5.0588 & 0 & 0 & 0\\
6.0588 & 0 & 0 & 0\\
6.0588 & 0 & 0 & 0\\
6.0588 & 0 & 0 & 0\\
6.0588 & 0 & 0 & 0\\
8.0588 & 0 & 0 & 0\\
8.0588 & 0 & 0 & 0\\
8.0588 & 0 & 0 & 0\\
\end{tabular}
\end{table}

\begin{table}[H]
\centering \protect\caption{$I=6$}
\begin{tabular}{r|rrrr}
E$\textbackslash J_B$ & 9 & 7 & 5 & 3\\
\toprule
1.0588 & 1.00003 & 0.00119549 & -0.00305693 & 0.0110501\\
2.8527 & -0.00220375 & 0.242714 & 0.805849 & 0.588035\\
3.0184 & -0.00390295 & 0.715143 & -0.523519 & 0.463442\\
3.1465 & -0.00375281 & -0.65544 & -0.276701 & 0.662776\\
3.8532 & 0 & 0 & 0 & 0\\
4.0549 & 0 & 0 & 0 & 0\\
4.1357 & 0 & 0 & 0 & 0\\
5.0588 & 0 & 0 & 0 & 0\\
5.0588 & 0 & 0 & 0 & 0\\
5.0588 & 0 & 0 & 0 & 0\\
6.0588 & 0 & 0 & 0 & 0\\
6.0588 & 0 & 0 & 0 & 0\\
6.0588 & 0 & 0 & 0 & 0\\
6.0588 & 0 & 0 & 0 & 0\\
8.0588 & 0 & 0 & 0 & 0\\
8.0588 & 0 & 0 & 0 & 0\\
8.0588 & 0 & 0 & 0 & 0\\
\end{tabular}
\end{table}

\begin{table}[H]
\centering \protect\caption{$I=8$}
\begin{tabular}{r|rrrrr}
E$\textbackslash J_B$ & 9 & 7 & 5 & 3 & 1\\
\toprule
1.0571 & 0.999799 & 0.00249462 & 0.0103254 & -0.0147855 & 0.0332888\\
2.0589 & -0.0037983 & 0.0689463 & 0.460992 & 0.850713 & 0.562402\\
2.8436 & -0.0155564 & 0.725503 & 0.149803 & -0.399347 & 0.568632\\
3.0588 & 0 & 0 & 0 & 0 & 0\\
3.1677 & -0.0112242 & -0.633338 & 0.531456 & -0.340893 & 0.34615\\
3.5299 & 0.00485083 & 0.260426 & 0.694712 & -0.0172994 & -0.489146\\
3.8645 & 0 & 0 & 0 & 0 & 0\\
4.0697 & 0 & 0 & 0 & 0 & 0\\
4.5287 & 0 & 0 & 0 & 0 & 0\\
5.0588 & 0 & 0 & 0 & 0 & 0\\
5.0588 & 0 & 0 & 0 & 0 & 0\\
6.0588 & 0 & 0 & 0 & 0 & 0\\
6.0588 & 0 & 0 & 0 & 0 & 0\\
8.0588 & 0 & 0 & 0 & 0 & 0\\
8.0588 & 0 & 0 & 0 & 0 & 0\\
\end{tabular}
\end{table}

\begin{table}[H]
\centering \protect\caption{$I=10$}
\begin{tabular}{r|rrrrr}
E$\textbackslash J_B$ & 9 & 7 & 5 & 3 & 1\\
\toprule
1.0464 & 0.997851 & 0.0321553 & 0.00357842 & 0.0887311 & -0.11316\\
2.0626 & -0.0423012 & 0.396662 & 0.860318 & -0.141699 & -0.721587\\
2.8207 & -0.0433641 & 0.674428 & -0.505476 & 0.761326 & -0.440961\\
3.0601 & 0 & 0 & 0 & 0 & 0\\
3.539 & -0.0259723 & -0.621868 & -0.0658534 & 0.626445 & -0.521579\\
4.0217 & 0 & 0 & 0 & 0 & 0\\
4.5109 & 0 & 0 & 0 & 0 & 0\\
5.0588 & 0 & 0 & 0 & 0 & 0\\
6.0588 & 0 & 0 & 0 & 0 & 0\\
8.0588 & 0 & 0 & 0 & 0 & 0\\
\end{tabular}
\end{table}

Overlaps for even $I$ and even $J_B$ ($T=1$):

\begin{table}[H]
\centering\protect\caption{$I=2$}
\begin{tabular}{r|r}
E$\textbackslash J_B$ & 8\\
\toprule
1.0589 & 0\\
3.0558 & 0\\
4.058 & 0.999981\\
5.0588 & 0\\
5.0588 & 0\\
5.0588 & 0\\
5.0588 & 0\\
6.0588 & 0\\
6.0588 & 0\\
6.0588 & 0\\
8.0588 & 0\\
8.0588 & 0\\
\end{tabular}
\end{table}

\begin{table}[H]
\centering\protect\caption{$I=4$}
\begin{tabular}{r|rr}
E$\textbackslash J_B$ & 8 & 6\\
\toprule
1.0588 & 0 & 0\\
3.028 & 0 & 0\\
3.0715 & 0 & 0\\
4.0308 & 0.160936 & 0.987333\\
4.0605 & -0.986952 & 0.158686\\
5.0588 & 0 & 0\\
5.0588 & 0 & 0\\
5.0588 & 0 & 0\\
5.0588 & 0 & 0\\
6.0588 & 0 & 0\\
6.0588 & 0 & 0\\
6.0588 & 0 & 0\\
6.0588 & 0 & 0\\
8.0588 & 0 & 0\\
8.0588 & 0 & 0\\
8.0588 & 0 & 0\\
\end{tabular}
\end{table}

\begin{table}[H]
\centering\protect\caption{$I=6$}
\begin{tabular}{r|rrr}
E$\textbackslash J_B$ & 8 & 6 & 4\\
\toprule
1.0588 & 0 & 0 & 0\\
2.8527 & 0 & 0 & 0\\
3.0184 & 0 & 0 & 0\\
3.1465 & 0 & 0 & 0\\
3.8532 & 0.0575001 & 0.530782 & 0.875848\\
4.0549 & 0.979765 & -0.19661 & 0.0512894\\
4.1357 & -0.191929 & -0.824416 & 0.479818\\
5.0588 & 0 & 0 & 0\\
5.0588 & 0 & 0 & 0\\
5.0588 & 0 & 0 & 0\\
6.0588 & 0 & 0 & 0\\
6.0588 & 0 & 0 & 0\\
6.0588 & 0 & 0 & 0\\
6.0588 & 0 & 0 & 0\\
8.0588 & 0 & 0 & 0\\
8.0588 & 0 & 0 & 0\\
8.0588 & 0 & 0 & 0\\
\end{tabular}
\end{table}

\begin{table}[H]
\centering\protect\caption{$I=8$}
\begin{tabular}{r|rrrr}
E$\textbackslash J_B$ & 8 & 6 & 4 & 2\\
\toprule
1.0571 & 0 & 0 & 0 & 0\\
2.0589 & 0 & 0 & 0 & 0\\
2.8436 & 0 & 0 & 0 & 0\\
3.0588 & 0.0163788 & 0.201778 & 0.702887 & 0.854136\\
3.1677 & 0 & 0 & 0 & 0\\
3.5299 & 0 & 0 & 0 & 0\\
3.8645 & 0.242862 & 0.800773 & -0.534145 & 0.252455\\
4.0697 & -0.967177 & 0.234518 & -0.068364 & 0.022665\\
4.5287 & -0.0726641 & -0.512835 & -0.464627 & 0.454166\\
5.0588 & 0 & 0 & 0 & 0\\
5.0588 & 0 & 0 & 0 & 0\\
6.0588 & 0 & 0 & 0 & 0\\
6.0588 & 0 & 0 & 0 & 0\\
8.0588 & 0 & 0 & 0 & 0\\
8.0588 & 0 & 0 & 0 & 0\\
\end{tabular}
\end{table}

\begin{table}[H]
\centering\protect\caption{$I=10$}
\begin{tabular}{r|rrrr}
E$\textbackslash J_B$ & 8 & 6 & 4 & 2\\
\toprule
1.0464 & 0 & 0 & 0 & 0\\
2.0626 & 0 & 0 & 0 & 0\\
2.8207 & 0 & 0 & 0 & 0\\
3.0601 & 0.126944 & 0.730409 & 0.788843 & -0.941105\\
3.539 & 0 & 0 & 0 & 0\\
4.0217 & 0.960518 & -0.279975 & 0.119751 & -0.0470487\\
4.5109 & 0.247442 & 0.623127 & -0.602827 & 0.334915\\
5.0588 & 0 & 0 & 0 & 0\\
6.0588 & 0 & 0 & 0 & 0\\
8.0588 & 0 & 0 & 0 & 0\\
\end{tabular}
\end{table}

Overlaps for odd $I$ and odd $J_B$ ($T=1$):

\begin{table}[H]
\centering \protect\caption{$I=3$}
\begin{tabular}{r|rr}
E$\textbackslash J_B$ & 8 & 6\\
\toprule
3.0658 & 0 & 0\\
4.0556 & 0.886512 & -0.465285\\
4.0694 & -0.462764 & -0.885217\\
5.0588 & 0 & 0\\
6.0588 & 0 & 0\\
6.0588 & 0 & 0\\
6.0588 & 0 & 0\\
6.0588 & 0 & 0\\
6.0588 & 0 & 0\\
8.0588 & 0 & 0\\
\end{tabular}
\end{table}

\begin{table}[H]
\centering \protect\caption{$I=5$}
\begin{tabular}{r|rrr}
E$\textbackslash J_B$ & 8 & 6 & 4\\
\toprule
3.0394 & 0 & 0 & 0\\
3.1403 & 0 & 0 & 0\\
4.0228 & 0.521481 & 0.528141 & -0.690876\\
4.0696 & -0.838204 & 0.461911 & -0.288948\\
4.1421 & -0.159705 & -0.712513 & -0.662573\\
5.0588 & 0 & 0 & 0\\
5.0588 & 0 & 0 & 0\\
6.0588 & 0 & 0 & 0\\
6.0588 & 0 & 0 & 0\\
6.0588 & 0 & 0 & 0\\
6.0588 & 0 & 0 & 0\\
6.0588 & 0 & 0 & 0\\
8.0588 & 0 & 0 & 0\\
\end{tabular}
\end{table}

\begin{table}[H]
\centering\protect\caption{$I=7$}
\begin{tabular}{r|rrrr}
E$\textbackslash J_B$ & 8 & 6 & 4 & 2\\
\toprule
2.8558 & 0 & 0 & 0 & 0\\
3.1128 & 0 & 0 & 0 & 0\\
3.5294 & 0 & 0 & 0 & 0\\
3.8495 & 0.242949 & 0.74858 & 0.120268 & -0.685731\\
4.0244 & 0.800366 & -0.487776 & 0.314524 & -0.224686\\
4.1601 & -0.547016 & -0.360402 & 0.533307 & -0.535697\\
4.5295 & -0.0329213 & -0.267937 & -0.775985 & -0.438508\\
5.0588 & 0 & 0 & 0 & 0\\
6.0588 & 0 & 0 & 0 & 0\\
6.0588 & 0 & 0 & 0 & 0\\
6.0588 & 0 & 0 & 0 & 0\\
6.0588 & 0 & 0 & 0 & 0\\
8.0588 & 0 & 0 & 0 & 0\\
\end{tabular}
\end{table}

\begin{table}[H]
\centering\protect\caption{$I=9$}
\begin{tabular}{r|rrrr}
E$\textbackslash J_B$ & 8 & 6 & 4 & 2\\
\toprule
2.0588 & 0 & 0 & 0 & 0\\
2.907 & 0 & 0 & 0 & 0\\
3.0586 & -0.063442 & -0.474189 & -0.854903 & -0.000975866\\
3.5251 & 0 & 0 & 0 & 0\\
3.8098 & 0.641427 & 0.30965 & -0.434754 & 0.753832\\
4.1659 & -0.727682 & 0.514185 & -0.282222 & 0.341678\\
4.5362 & -0.234441 & -0.644132 & 0.021184 & 0.561164\\
6.0588 & 0 & 0 & 0 & 0\\
6.0588 & 0 & 0 & 0 & 0\\
8.0588 & 0 & 0 & 0 & 0\\
\end{tabular}
\end{table}

Overlaps for odd $I$ and odd $J_B$ ($T=0$):

\begin{table}[H]
\centering\protect\caption{$I=3$}
\begin{tabular}{r|r}
E$\textbackslash J_B$ & 9\\
\toprule
3.0658 & -1.00006\\
4.0556 & 0\\
4.0694 & 0\\
5.0588 & 0\\
6.0588 & 0\\
6.0588 & 0\\
6.0588 & 0\\
6.0588 & 0\\
6.0588 & 0\\
8.0588 & 0\\
\end{tabular}
\end{table}

\begin{table}[H]
\centering\protect\caption{$I=5$}
\begin{tabular}{r|rr}
\toprule
E$\textbackslash J_B$ & 9 & 7\\
3.0394 & -0.933845 & 0.37385\\
3.1403 & 0.357561 & 0.927537\\
4.0228 & 0 & 0\\
4.0696 & 0 & 0\\
4.1421 & 0 & 0\\
5.0588 & 0 & 0\\
5.0588 & 0 & 0\\
6.0588 & 0 & 0\\
6.0588 & 0 & 0\\
6.0588 & 0 & 0\\
6.0588 & 0 & 0\\
6.0588 & 0 & 0\\
8.0588 & 0 & 0\\
\end{tabular}
\end{table}

\begin{table}[H]
\centering\protect\caption{$I=7$}
\begin{tabular}{r|rrr}
E$\textbackslash J_B$ & 9 & 7 & 5\\
\toprule
2.8558 & -0.472944 & -0.73643 & 0.636574\\
3.1128 & 0.874802 & -0.431482 & 0.189249\\
3.5294 & 0.105405 & 0.521038 & 0.747663\\
3.8495 & 0 & 0 & 0\\
4.0244 & 0 & 0 & 0\\
4.1601 & 0 & 0 & 0\\
4.5295 & 0 & 0 & 0\\
5.0588 & 0 & 0 & 0\\
6.0588 & 0 & 0 & 0\\
6.0588 & 0 & 0 & 0\\
6.0588 & 0 & 0 & 0\\
6.0588 & 0 & 0 & 0\\
8.0588 & 0 & 0 & 0\\
\end{tabular}
\end{table}

\begin{table}[H]
\centering\protect\caption{$I=9$}
\begin{tabular}{r|rrrr}
E$\textbackslash J_B$ & 9 & 7 & 5 & 3\\
\toprule
2.0588 & -0.194963 & -0.741037 & -0.769821 & 0.929465\\
2.907 & -0.877203 & 0.463008 & -0.267855 & 0.124265\\
3.0586 & 0 & 0 & 0 & 0\\
3.5251 & 0.43861 & 0.486313 & -0.579331 & 0.347367\\
3.8098 & 0 & 0 & 0 & 0\\
4.1659 & 0 & 0 & 0 & 0\\
4.5362 & 0 & 0 & 0 & 0\\
6.0588 & 0 & 0 & 0 & 0\\
6.0588 & 0 & 0 & 0 & 0\\
8.0588 & 0 & 0 & 0 & 0\\
\end{tabular}
\end{table}

}
\end{document}